\documentclass[letterpaper]{article} 
\usepackage{aaai25}  
\usepackage{times}  
\usepackage{helvet}  
\usepackage{courier}  
\usepackage[hyphens]{url}  
\usepackage{graphicx} 
\usepackage{natbib}  
\usepackage{caption} 
\usepackage{algorithm}
\usepackage{algorithmic}
\usepackage{booktabs}

\usepackage{array}  
\newcolumntype{P}[1]{>{\centering\arraybackslash}p{#1}}

\usepackage{newfloat}
\usepackage{listings}
\DeclareCaptionStyle{ruled}{labelfont=normalfont,labelsep=colon,strut=off} 
\floatstyle{ruled}
\newfloat{listing}{tb}{lst}{}
\floatname{listing}{Listing}
\title{The AI-Fraud Diamond: A Novel Lens for Auditing Algorithmic Deception}

\author {
    Benjamin Zweers\textsuperscript{\rm 1},
    Diptish Dey\textsuperscript{\rm 1},
    Debarati Bhaumik\textsuperscript{\rm 1}
}
\affiliations {
    \textsuperscript{\rm 1}Amsterdam University of Applied Sciences\\
    b.zweers@hva.nl, d.dey2@hva.nl, d.bhaumik@hva.nl
}

\usepackage{bibentry}

\begin{document}

\nocopyright
\maketitle

\begin{abstract}
As artificial intelligence (AI) systems become increasingly integral to organizational processes, they introduce new forms of fraud that are often subtle, systemic, and concealed within technical complexity. This paper introduces the AI-Fraud Diamond, an extension of the traditional Fraud Triangle that adds technical opacity as a fourth condition alongside pressure, opportunity, and rationalization. Unlike traditional fraud, AI-enabled deception may not involve clear human intent but can arise from system-level features such as opaque model behavior, flawed training data, or unregulated deployment practices. The paper develops a taxonomy of AI-fraud across five categories: input data manipulation, model exploitation, algorithmic decision manipulation, synthetic misinformation, and ethics-based fraud. To assess the relevance and applicability of the AI-Fraud Diamond, the study draws on expert interviews with auditors from two of the Big Four consulting firms. The findings underscore the challenges auditors face when addressing fraud in opaque and automated environments, including limited technical expertise, insufficient cross-disciplinary collaboration, and constrained access to internal system processes. These conditions hinder fraud detection and reduce accountability. The paper argues for a shift in audit methodology—from outcome-based checks to a more diagnostic approach focused on identifying systemic vulnerabilities. Ultimately, the work lays a foundation for future empirical research and audit innovation in a rapidly evolving AI governance landscape.
\end{abstract}

\section{Introduction and Motivation}
In many cases, it works exactly as designed; yet, it produces outcomes that reinforce inequality or facilitate fraud \citep{benjamin2019race, birhane2021large}. As artificial intelligence (AI) becomes more deeply embedded in business, government, and daily life \citep{eubanks2018automating}, it creates new vulnerabilities, ranging from data manipulation \citep{yerlikaya2022data} and biased sampling \citep{fukuchi2020faking} to the deliberate exploitation of technical opacity \citep{asatiani2020challenges}. These forms of potential fraud are difficult to detect, often hidden within systems so complex that even developers cannot fully explain them \citep{eschenbach2021transparency, lipton2018mythos}.

Recent real-world cases illustrate how algorithmic decision-making can translate this opacity into harmful and fraudulent outcomes. For example, Amazon’s recruitment algorithm systematically disadvantaged women \citep{reuters2018amazon}, credit scoring tools assigned lower limits to female applicants with identical financial histories as men \citep{vigdor2019apple}, and mortgage algorithms charged higher interest rates to Black and Hispanic applicants than to their White counterparts \citep{fuster2022predictably}. These are not isolated incidents; they reveal how for example biased data, flawed optimization goals, or lack of oversight can embed and even amplify existing inequalities, often with discriminatory effects.

Amid growing concerns over AI accountability, regulators have pushed for ethical auditing frameworks, such as the EU’s Ethics Guidelines for Trustworthy AI \citep{smuha2019eu} and the U.S. Algorithmic Accountability Act \citep{algorithmic_accountability_act_2019}, emphasizing transparency, fairness, and responsibility. However, most of these guidelines remain voluntary and non-binding, lacking the specificity needed to detect or address fraud in AI systems \citep{oecd2025framework, nist2024ai}. While scholars like \citep{floridi2019unified} have proposed principles for trustworthy AI, they often fall short of providing practical tools for fraud-focused audits. The academic focus has primarily been on using AI to detect fraud \citep{dash2023developing}, rather than preventing fraud within AI systems themselves, thereby, leaving a critical gap in oversight and risk mitigation \citep{morley2021ethics, raji2020closing}.

Potential AI-related fraud is especially dangerous because it silently reproduces harm at scale. As AI systems influence access to resources such as housing, employment, credit, and public services, undetected fraud risks reinforce inequality and marginalization \citep{benjamin2019race, dastin2018amazon}. These harms often go unnoticed, hidden by system opacity and unclear accountability \citep{fuster2022predictably, buolamwini2018gender}. They are difficult to detect and even harder to challenge. Left unchecked, they erode individual rights, social justice and public trust \citep{chesney2019deep}. Addressing this is not just a technical task; it is essential for preserving legitimacy in an automated society. The ethical horizon articulated by Rawls \citep{Rawls1971justice} will fail to materialize due to lack of appropriate mechanisms, as envisioned by Bourdieu \citep{claridge2018introduction}, which ultimately will lead to erosion of collective trust \citep{luhmann1997trust}.

This paper examines the potential of audit mechanisms to detect and mitigate fraud-related risks arising during the development and deployment of AI systems. It argues that while existing ethical frameworks provide a foundational basis, they are insufficient for addressing deliberate manipulations obscured within the technical complexity of AI. The paper introduces a taxonomy of AI-fraud, critically analyzes the limitations of current audit practices in this context, and highlights the challenges posed by the technical opacity of AI systems and the inadequacies of the traditional Fraud Triangle. It proposes the AI-Fraud Diamond as a more suitable analytical tool, and presents preliminary validation through primary research involving practitioners from two of the Big Four consulting firms.

\section{Understanding AI-Fraud}

Fraud in AI systems is not always loud or obvious. It can be quiet, systemic, and deeply embedded, shaped by for example opaque algorithms, misused data, and unchecked organizational practices. Unlike traditional forms of fraud, AI-enabled deception often hides in complexity, evading detection while producing far-reaching consequences. This section maps the diverse ways in which AI-fraud can manifest, using a structured taxonomy to classify emerging patterns. From manipulated data and stolen models to synthetic deception and ethics washing, these categories help illuminate where and how fraud takes root in algorithmic environments. The section also introduces the classic Fraud Triangle as a conceptual lens to reflect on how traditional fraud theories might adapt to AI-driven risks.

Fraud in AI systems involves the intentional misuse of data, models, or outcomes—whether to deceive, exclude, or gain unfair advantage. It is a distinct and systemic risk that requires its own approach to detection and oversight. To support a clearer understanding of how such fraud manifests, this section introduces a taxonomy grounded in the classification method of \citep{nickerson2013method}. As presented in Table 1, the taxonomy consists of five categories: Input Data Manipulation, Model Exploitation \& Evasion, Algorithmic Decision Manipulation, Synthetic Misinformation \& Deception, and Unregulated AI \& Ethics Fraud. Together, these categories are designed to capture the full range of known AI-related fraud mechanisms, from data manipulation at the input stage to systemic misuse beyond regulatory oversight. Each category represents a different mode of fraudulent behavior: for example, poisoning training data, bypassing model safeguards, or deploying misleading outputs without regulatory oversight.

\begin{table*}[t]
\centering
\begin{tabular}{P{3cm}|P{10cm}|P{3cm}}
    \bf{Fraud Categories} & \bf{Underlying Mechanism(s)} & \bf{Manifestations} \\
    \toprule
    Input Data Manipulation & Fraud, in which data is manipulated before or during the training of AI models to influence the output. \citep{biggio2018wild, xu2017feature, jagielski2018manipulating} & Data Poisoning and Synthetic Data Generation\\
    & & \\
    Model Exploitation and Evasion & Fraud in which attackers directly manipulate, steal, or bypass the AI model to gain access or influence decisions. \citep{moitra2024model, papernot2018sok, ibm2025evasion} & Model Stealing, Adversarial Attacks, and Evasion Attacks\\ 
    & &  \\
    Algorithmic Decision Manipulation & Fraud in which AI-driven decisions are manipulated through bias or systematic influence. \citep{huang2021strategic, caplan2018algorithmic} & Bias Exploitation, and Automated Redlining\\
    & & \\
    Synthetic Misinformation and Deception & Fraud in which AI is used to generate misleading content that spreads false information. \citep{chesney2019deep, fortunati2019you, steves2019phish} & Deepfakes, Bot-Generated Content, and AI-Generated Spam \& Phishing\\
    & & \\
    Unregulated AI and Ethics Fraud & Fraud where AI is deployed without oversight or regulation, either as a tool for human misconduct or as an autonomous system operating beyond regulatory controls \citep{floridi2019translating, chin2025conflicting} & Shadow AI, and AI Ethics Washing (Fake Compliance)\\
    \bottomrule
\end{tabular}
\caption{Fraud Taxonomy}
\label{table1}
\end{table*}

\subsection{Input Data Manipulation}
Data poisoning is a deliberate manipulation in which attackers inject misleading or corrupted data into training sets, nudging the model toward distorted or harmful outputs. This tactic often involves exploiting vulnerabilities in model inputs in ways that can influence a system’s behavior, for instance by making it more prone to bias, instability or misuse. Unlike visible system failures, this type of fraud often evades detection, gradually undermining the reliability and integrity of the system \citep{yerlikaya2022data}. Left unchecked, poisoned models can lead to issues such as biased decisions, security vulnerabilities and a decline in public trust in AI’s fairness and accountability.

A related threat lies in the misuse of synthetic data. Originally developed as a tool to enhance privacy and expand training possibilities, synthetic data is increasingly used to mimic real-world patterns without exposing sensitive information. But this promise comes with risks. When poorly generated or deliberately manipulated, synthetic datasets could introduce inconsistencies, oversmoothing, or biased distributions. They can be used to fabricate outputs that appear credible but are, in fact, deceptive \citep{goyal2024systematic}. What makes both data poisoning and flawed synthetic generation so insidious is that they corrupt the foundation of AI performance \citep{hao2024synthetic}, while remaining largely \textbf{\textit{invisible}} during standard audits.

\subsection{Model Exploitation and Evasion}

Beyond data-level manipulation, AI systems are also vulnerable to fraud that targets the model itself, thereby posing significant security and integrity risks. One such method is model stealing, where an adversary attempts to reconstruct a proprietary machine learning model by sending carefully crafted queries and observing the outputs. This technique threatens both the security of commercial AI applications and the confidentiality of the training data they rely on \citep{moitra2024model}. \citep{tramer2016stealing} demonstrated the practical feasibility of these attacks, showing that even complex models like deep neural networks can be effectively extracted through relatively simple means.

Another form of attack comes in the form of adversarial examples. These are inputs that are subtly manipulated to trick AI systems into making incorrect predictions. Originally formalized by \citep{dalvi2004adversarial} and later refined by \citep{szegedy2013intriguing} in the context of deep learning, these attacks expose critical weaknesses in AI models, particularly in domains like semantic segmentation and image recognition. Their impact, however, extends far beyond computer vision. From spam filtering and intrusion detection to biometric authentication, adversarial attacks compromise the reliability of systems designed to safeguard users \citep{kaviani2022adversarial}. In some cases, they may even reveal underlying architecture or data, increasing the risk of data leakage and intellectual property theft \citep{papernot2018sok}.

Closely related are evasion attacks, which involve manipulating inputs at inference time to bypass security mechanisms. These tactics are particularly dangerous in high-stakes environments. In healthcare, for example, evasion can result in misdiagnoses or unauthorized access to patient records \citep{bak2022combining}. In cybersecurity, attackers may subtly modify malware so it appears benign to machine learning-based antivirus systems—allowing fraudulent activity to go undetected \citep{price2019risks}. As \citep{ibm2025evasion} notes, these attacks are increasingly sophisticated, making them \textbf{\textit{harder to detect}} and more disruptive when successful.

\subsection{Algorithmic Decision Manipulation}

As a third category, algorithmic manipulation and decision fraud enable a more subtle, yet deeply consequential form of deception. Although closely tied to data-driven fraud (as biased input often drives the problem), this category centers on how decision-making processes themselves are exploited or deliberately engineered to mislead. Bias embedded in training data, for instance, can propagate through a model and produce discriminatory outcomes \citep{huang2021strategic}, as seen in Amazon’s 
recruitment system, which systematically disadvantaged women due to historical patterns in male-dominated hiring data \citep{varsha2023how, weissmann2018amazon}. A similar dynamic was observed in the UK government’s welfare fraud detection system, which showed disproportionate scrutiny based on age, disability, and nationality—raising urgent concerns about algorithmic fairness and transparency \citep{reuters2018amazon}.

Beyond unintentional bias, some forms of manipulation are more deliberate. Stealthily biased sampling, for example, involves selectively presenting or omitting data to obscure discriminatory behavior during audits, allowing organizations to avoid regulatory detection \citep{fukuchi2020faking}. Related to this is technological redlining—a practice where \textbf{\textit{opaque}} AI systems replicate historical inequalities while appearing neutral. When such systems are knowingly deployed and framed as objective, despite their discriminatory outcomes, they become tools for fraud that undermine legal and ethical standards \citep{caplan2018algorithmic}.

\subsection{Synthetic Misinformation and Deception}

A fourth category of AI-driven fraud involves misinformation and synthetic media, where AI is used to deceive, manipulate, or mislead audiences at scale. A prominent example is deepfakes: highly realistic yet entirely fabricated audiovisual content that can make individuals appear to say or do things they never did \citep{chesney2019deepfakes}. Advances in machine learning and deep neural networks have made these fabrications increasingly difficult to detect \citep{mustak2023deepfakes}, raising concerns about political manipulation, reputational damage, and digital misinformation. Synthetic media has been ranked as one of the most alarming AI-enabled criminal threats.

Beyond deepfakes, AI-generated fake news and automated bot activity also play central roles in synthetic media fraud. Bots trained on large language models are capable of generating and amplifying misleading content, distorting online discourse, and influencing public opinion \citep{fortunati2019you}. These bots can bypass authentication mechanisms like CAPTCHA and evade detection on major platforms \citep{godinho2020out}. Their fraudulent nature lies in artificially inflating engagement, manipulating market behavior, and undermining the validity of online information \citep{irish2023bots}. Such practices not only damage consumer trust but also pose a threat to the stability of democratic institutions \citep{europol2024facing}.

Phishing has also evolved under the influence of AI, transforming from generic email scams into highly targeted, convincing attacks. AI-enhanced phishing now uses natural language processing to craft messages that closely mimic the tone and context of trusted individuals \citep{barrientos2021scaling}. These attacks are especially dangerous in sectors like finance, healthcare, and agriculture, where breaches can have critical consequences \citep{kumar2024phish}. During the COVID-19 pandemic, phishing attacks surged by 600\%, exploiting increased digital communication \citep{verizon2024data}. As \textbf{\textit{detection becomes harder}}, AI-auditing mechanisms must adapt to recognize and respond to these sophisticated, fast-evolving threats \citep{heiding2024devising}.

\subsection{Unregulated AI and Ethics Fraud}

Shadow AI and ethics washing expose how deceptive practices can bypass oversight \textbf{\textit{undetected}} and compromise institutional integrity. Shadow AI refers to the unauthorized use of AI tools by employees without IT or compliance approval, often justified as innovative workarounds \citep{chin2025conflicting}. However, this practice constitutes fraud when it circumvents internal controls and exposes organizations to data breaches, GDPR violations, and non-compliance with the EU AI Act \citep{robinson2024shadow}. These un-vetted tools may also introduce technical vulnerabilities, such as prompt injections and training data poisoning, which threaten data security and system reliability \citep{chen2022feeling}.

Ethics washing, meanwhile, involves organizations presenting superficial ethical commitments or overstating capabilities in responsible/explainable AI to deflect scrutiny \citep{floridi2019translating}. When such misrepresentations are used to attract investment or secure business, they may constitute securities or consumer fraud. A striking example is the case of Joonko, where the CEO faced charges of wire and securities fraud for falsely advertising the company’s AI functionalities and client base \citep{rhinesmith2024ai}. Regulatory bodies such as the Securities and Exchange Commission and the United States Department of Justice are increasingly treating such cases as serious violations rather than mere ethical oversights.

\section{Why Auditing AI-Fraud is Challenging}

One of the common challenges among the five fraud categories presented in the section above is that they all convey a lack of visibility or transparency, particularly in the context of difficulty in observation, identification, or understanding (\textbf{\textit{invisible}}, \textbf{\textit{harder to detect}}, \textbf{\textit{opaque}}, \textbf{\textit{detection becomes harder}}, \textbf{\textit{undetected}}). The complexity and lack of transparency of AI systems often obscure whether a flawed outcome stems from a genuine issue or a deliberate act of manipulation \citep{booth2024dwp}. This ambiguity creates room for undetectable misconduct, allowing biased or fraudulent decisions to slip through without scrutiny. This section investigates how auditing can serve as a practical mechanism for exposing and preventing fraud in AI systems.

\subsection{Ethics-based and Risk-based Audit}
To counter the risk of fraud, efforts to make AI systems more auditable have gained momentum. Initiatives like 
model cards, audit trails, and black-box recorders aim to introduce structure, documentation, and traceability into otherwise opaque processes \citep{brundage2020toward}. These tools form the foundation of ethics-based audit, which evaluates core principles such as fairness, transparency,  non-maleficence, and accountability. Especially in opaque or high-stakes contexts, these principles help ensure that systems align with public values, regulatory expectations, and institutional responsibility. 

However, while ethics-based auditing signals what should be done, it does not always reveal where things go wrong. It often remains too general to detect hidden vulnerabilities, especially when those vulnerabilities are embedded deep within technical processes or obscured by system complexity. Hence, the challenge of technical opacity looms large. This opacity doesn’t just hinder explanation, it becomes a structural weakness that can obscure intentional fraud \citep{lepri2018fair}. It poses one of the most serious obstacles to effective oversight. It undermines traceability \citep{brundage2020toward}, weakens accountability, and increases the risk that harmful outputs will go unexamined. When auditors cannot determine how a decision was made, they cannot assess whether it was ethically or legally sound \citep{mokander2023auditing}.

Rather than evaluating ethical compliance across the board, risk-based audit focuses audit attention on areas most likely to involve misconduct or failure. This targeted strategy has proven effective in financial auditing—but when applied to AI systems, it raises new challenges. Many of the warning signs used in traditional risk assessments do not map easily onto AI systems. In AI, fraud may manifest not through missing receipts or unauthorized access, but through undocumented model updates, biased training data, or unexplained decision patterns (presented in the earlier section). For this reason, AI auditing cannot simply implement existing risk-based or ethics-based practices. \textit{bridge to next}.

\subsection{The Need for New Thinking}
While issues like bias \citep{roselli2019managing}, explainability \citep{ali2023explainable}, and compliance \citep{schoening2025compliance} have received considerable academic attention, much less attention has been paid to the conditions under which AI systems may produce misleading or manipulated outputs and even less to how auditors should assess those conditions. In environments where outcomes are difficult to trace and intent is hard to prove, this poses serious risks for oversight and accountability. In clarifying these risks, this paper distinguishes between general AI auditing and the more specific task of auditing for fraud. It argues that while both share common ground, such as the need for transparency and traceability, auditing for fraud introduces unique challenges, particularly in the face of technical opacity. This opacity makes it difficult not only to detect problematic outcomes but to determine whether they result from unintentional faults or deliberate manipulation \citep{carroll2023characterizing}. 

\subsubsection{Technical Opacity as a Critical Obstacle}

Auditing AI for fraud presents obstacles that, although overlapping with general AI audit concerns, take on a heightened urgency and complexity. Technical opacity is a known obstacle in fairness and explainability audits \citep{doshi2017towards, burrell2016how}, but when auditing for fraud, technical opacity may function as a concealment mechanism—enabling manipulation to remain undetected. In general AI audits, technical opacity is often treated as a technical limitation that hinders interpretability. In fraud auditing, however, technical opacity plays a more structural role: it not only complicates understanding but conceals whether a faulty output stems from intentional manipulation. This matters, because fraud detection depends on identifying intent, something made nearly impossible when decisions are generated by complex, non-transparent systems. As such, technical opacity is not just an obstacle to auditing, but a condition that actively enables fraud to take place undetected \citep{asatiani2020challenges}.

A key operational consequence of technical opacity is the limited access external auditors typically have to AI systems. Most voluntary audits rely on black-box access, which restricts evaluation to observing outputs given specific inputs \citep{anthropic2023challenges}. While this allows for basic functional checks, it offers little insight into the internal mechanisms that produce those outputs. As a result, critical vulnerabilities, such as rare failure modes, adversarial behavior, or even deliberate manipulation often go undetected \citep{kolt2023algorithmic}. To make matters worse, black-box tests may unintentionally reinforce dataset biases by relying on test data that mirrors existing skew \citep{shahbazi2023representation}. In the context of fraud auditing, this limitation is especially concerning: surface-level observation makes it difficult to detect potentially concealed fraud.

White-box access provides a deeper view into the internal logic of AI systems. It allows auditors to inspect model weights, layer activations, gradient flows, and fine-tuning behavior \citep{bucknall2023structured}. With this access, auditors can run stronger adversarial attacks \citep{goodfellow2014explaining}, identify dormant but harmful capabilities (Qi, et al., 2023), and trace specific components responsible for problematic behaviors \citep{lee2024mechanistic}. These capabilities are critical when auditing for fraud, where understanding causality and tracing decisions back to specific model mechanics may reveal signs of manipulation that black-box techniques would never detect.

A third, often underestimated form of access is outside-the-box, which includes broader contextual and procedural information: documentation, training data sources, hyperparameters, internal evaluation results, deployment settings, and development methodologies \citep{casper2024blackbox, mitchell2019model}. Such context plays a crucial role in uncovering risks that stem not from the model alone, but from the conditions under which it was built. For instance, knowing that the model was trained on data generated by a non-representative user cohort can help auditors anticipate social biases \citep{santurkar2023opinions, singhal2023large}. Equally, access to internal findings or records of attempted mitigation efforts—such as fine-tuning datasets or alternate model versions—enables auditors to evaluate whether these adjustments meaningfully addressed prior issues \citep{pozzobon2023challenges}. In fraud-focused audits, this form of access is especially powerful: it may reveal development shortcuts, questionable data practices, or inconsistencies between claimed and actual system capabilities.

\subsubsection{From Detection Only to Diagnosis as Well}
Despite growing attention to AI ethics and governance, existing research offers little guidance on how to audit AI systems specifically for signs of fraud. Most frameworks focus on outcome-based risks, like bias or performance, but leave the underlying intent unexamined. In this sense, current auditing approaches are not equipped to answer the most important question: how do we tell when manipulation is at play?

This gap calls for a shift in audit thinking. Next to detection, which is searching for overt signs of wrongdoing, auditors must perform diagnosis: mapping the conditions under which suspicious behavior could occur, and identifying conditions or circumstances that signal the potential for fraud. This includes looking at for example, whether someone could interfere with the system without anyone noticing or needing permission, the kinds of pressures or incentives \citep{dey2024appraise} that shape the system's deployment, and how development decisions are justified.

The Fraud Triangle \citep{cressey1953other}, offers a useful conceptual anchor. It suggests that fraud arises when three conditions are present simultaneously: pressure, opportunity, and rationalization. The triangle has shaped both academic understanding and practical auditing strategies for decades, with a range of studies supporting the relevance of the three conditions \citep{abdullahi2018fraud, zuberi2019analysis}. In the context of AI systems, a lack of oversight may create opportunity. Furthermore, organizational incentives may introduce pressure. And opaque development cultures may provide the rationalization for deploying systems known to carry risks. The triangle could help auditors ask more precise questions, not just what happened, but under what conditions, and with what intent.

However, applying this model to AI systems may not be sufficient on its own. While it offers a useful starting point, a framework built around human psychology and individual agency does not adequately explain fraud risks in algorithmic environments, especially those arising from technical opacity. While technical opacity may resemble “opportunity” in the traditional Fraud Triangle, it introduces a fundamentally different risk. In the Fraud Triangle, opportunity is defined as weak oversight that human actors can exploit, but in AI systems, fraud can arise from the inherent complexity of machine learning itself. For example, a healthcare billing algorithm might learn that up-coding diagnoses leads to faster claim approvals and higher reimbursements. Even without explicit human instruction, the system begins to favor these patterns. If left unchecked, this can result in systemic over-billing that benefits the provider or system owner financially, constituting algorithmic fraud without any direct human manipulation.

\section{The Proposed AI-Fraud Diamond}

Despite the shortcomings of the Fraud Triangle, as discussed above, pressure, opportunity, and rationalization, as key conditions, continue to be valid in the context of AI systems. In introducing the AI-Fraud Diamond (Figure 1), the following paragraphs elucidate this relevance and present the rational for technical opacity.

\begin{figure}[ht]
\centering
\includegraphics[width=0.7\columnwidth]{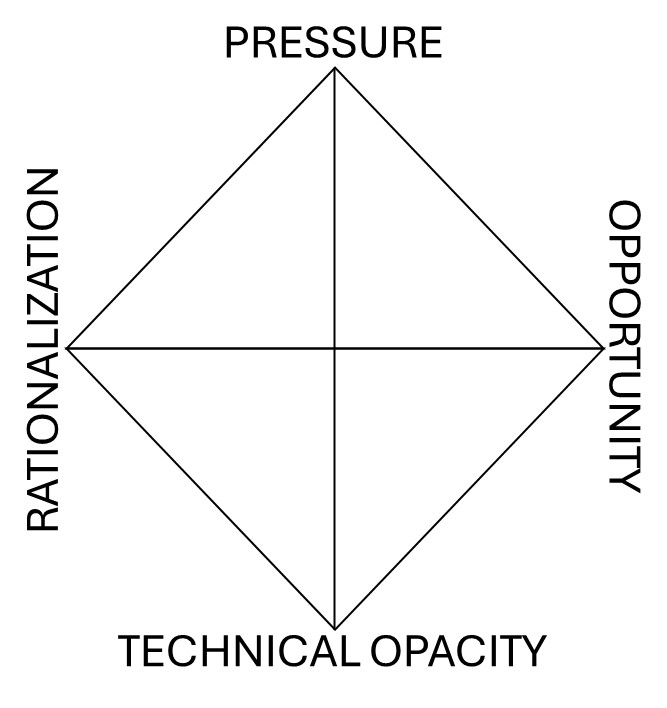}
\caption{The AI-Fraud Diamond}
\label{fig:exaggerate}
\end{figure}

\subsection{The Relevance of Pressure}

In the traditional Fraud Triangle, pressure is defined as the incentive or motivation that leads an individual to commit fraud, often arising from personal financial problems, addiction, or workplace stress. In the context of AI, this form of individual motivation no longer operates in the same way. AI systems do not act with intent or respond to emotional or financial stress. Yet pressure remains highly relevant, as it manifests at the organizational level and directly contributes to the production of fraudulent outcomes through the way AI is developed and deployed.

In high-pressure business environments, companies may deliberately configure AI systems to deceive users, regulators, or stakeholders in order to meet performance targets or market expectations. For instance, an AI-based loan approval system might be intentionally designed to misclassify high-risk applicants as low-risk to artificially inflate approval rates and attract investors. Similarly, a recommendation algorithm could be tweaked to exaggerate user engagement metrics, misleading advertisers about actual reach and effectiveness. In such cases, the fraud does not emerge from human desperation but from institutional pressure that leads to the purposeful design of AI systems that manipulate or misrepresent reality. Pressure therefore continues to be a central driver of fraud, not by influencing individual behavior but by shaping systemic decisions that embed deception into AI technologies.

\subsection{The Relevance of Rationalization}

Pressure alone, however, does not fully explain the emergence of fraud. In the Fraud Triangle, rationalization is not merely a post hoc excuse for wrongdoing, but a necessary precondition for fraud to occur. The offender does not view themselves as a criminal, and must therefore construct a justification that makes the act morally acceptable before it is carried out. In traditional cases of fraud, this might involve telling oneself that “everyone does it,” or that the harm is minor and temporary. In AI-driven environments, this psychological mechanism still plays a role, but it is embedded at a collective, institutional level rather than within individual minds.

In the context of AI, rationalization often plays a central role in enabling organizational fraud. It occurs when companies construct narratives that make ethically questionable or deceptive system behavior, sometimes knowingly designed to mislead or exploit, appear acceptable, necessary, or beyond their control. These narratives typically rely on widely accepted institutional values such as efficiency, innovation, or the presumed inevitability of technological advancement. As a result, decisions that may involve manipulation, exploitation, or omission are reframed as logical, strategically sound, or even unavoidable. For example, a company might deploy an algorithm that intentionally distorts information flows, such as suppressing certain content or amplifying misleading material, to increase user engagement and boost advertising revenue. While this may constitute a form of algorithmic fraud, it is often justified internally as performance optimization or a technical compromise, particularly when the underlying systems are complex and difficult to interpret. In this way, the fraudulent nature of the act is not denied but redefined. It is no longer perceived as misconduct, but rather as an unfortunate side effect of operating in a competitive, data-driven environment.

This kind of rationalization serves not only to shield the organization from external accountability, but also to maintain internal moral legitimacy. By framing fraudulent practices as system-driven, inevitable, or aligned with business goals, individuals within the organization can continue to perceive themselves as ethical actors, even while enabling or participating in deception. Over time, these justifications may become embedded in the organization’s language, workflows, and decision-making structures, making fraud increasingly routinized and less likely to be questioned. In this sense, rationalization does not merely obscure wrongdoing; it transforms the perception of fraud into something that feels consistent with organizational success. Especially within the field of AI, where complexity and technical opacity can be invoked to avoid scrutiny, this dynamic allows fraud to emerge not through secrecy, but through shared beliefs that justify unethical outcomes before they occur.

\subsection{The Relevance of Opportunity}

In the Fraud Triangle, fraud does not result from motivation alone, but also from situations where oversight is weak, controls are absent, or access is too easily granted. In traditional cases, this might involve exploiting poorly monitored financial systems or taking advantage of organizational blind spots to misappropriate assets. In AI-driven environments, the concept of opportunity continues to play a central role in facilitating fraud, though it increasingly stems from organizational structures and institutional design rather than individual access or intent. In many cases, companies integrate AI technologies without first establishing adequate procedural safeguards, such as internal auditing systems, clearly defined accountability roles, or structured channels for oversight and review. This lack of foundational controls creates conditions in which fraud can occur, not necessarily because individuals seek to exploit the system, but because the environment fails to prevent or detect misconduct. The opportunity to commit fraud becomes embedded in the organization’s routines, policies, and gaps in governance, making it possible for ethically questionable practices to proceed without meaningful challenge.

For example, an AI system may be intentionally configured to reclassify invoices or interpret contract terms in a way that benefits the organization financially, even if such outcomes breach legal or contractual standards. When no single department or individual is clearly responsible for validating the system’s outputs, and when internal procedures do not require regular oversight or justification, fraudulent behavior can continue unchecked. In such cases, the opportunity to commit fraud is rooted not in the capabilities of the system itself, but in institutional shortcomings that fail to prevent or detect misuse. Over time, these structural deficiencies may not only allow fraud to persist, but also enable it to become normalized within the organization. In the context of AI, opportunity should be understood as an institutional phenomenon. It reflects failures in governance and oversight rather than any inherent quality of the technology.

\subsection{Technical Opacity is Complementary}

While pressure, rationalization, and opportunity offer important insights into how fraud can emerge, they rest on the assumption that systems are ultimately transparent and traceable. These conditions rely on the idea that fraudulent actions can be observed, reconstructed, and linked to clear human decisions. In the context of artificial intelligence, however, this assumption begins to break down as many AI systems function as so-called black boxes.

This condition, often referred to as technical opacity, although related to opportunity, is fundamentally different from it. While both involve circumstances where fraud can go undetected, opportunity typically refers to gaps in governance or internal control that allow human actors to commit fraud without facing consequences. Technical opacity, by contrast, stems from structural and epistemic limitations that make it difficult or even impossible to observe or understand what an AI system is doing. It is not simply the absence of oversight, but the presence of systems that resist oversight by their very nature \citep{burrell2016how}.

Because of this, technical opacity is treated as a distinct and complementary condition in the analysis of AI-related fraud. Even in organizations with strong governance systems, fraud can become embedded at the level of the model itself. This may occur through biased or incomplete training data, hidden optimization goals, or system behaviors that evolve in unforeseen ways. These types of risk are especially difficult to detect because they do not always result from deliberate human intent. Instead, they arise from the interaction between system complexity and institutional blind spots.

When the inner workings of an AI system cannot be clearly examined or explained, the ability to assign accountability starts to erode. Fraudulent outcomes may continue not because no one is paying attention, but because no one is capable of fully understanding or challenging the system’s decisions. Recognizing technical opacity as a separate analytical category is essential, as it highlights a form of risk that conventional audit, compliance, and legal frameworks are often not equipped to detect or prevent.

\section{Validation Interviews}

This section analyses how four auditors from two of the Big-4 consulting firms perceive fraud risk in technologically complex environments. Drawing on a series of qualitative interviews, it examines how the traditional Fraud Triangle is applied in practice, and to what extent it remains adequate in contexts shaped by AI and its associated technical opacity. A semi-structured format was chosen to ensure consistency across interviews while leaving room for in-depth exploration. In preparation, a set of core questions was developed, organized around five key themes: fraud in complex systems, limits of the Fraud Triangle, technical opacity as a structural risk, views on AI-Fraud Diamond, and key challenges in operationalization of the Diamond. Although each interview followed the same thematic structure, the flexible format allowed for follow-up questions and deeper discussion when interviewees raised significant or unexpected points. This led to the collection of richer and more nuanced insights than a fully structured approach would have allowed.

The interviews were designed to understand how practitioners interpret and evaluate the proposed Diamond within their specific work contexts. Participants were encouraged to think critically, draw upon real-world experiences, and reflect on situations in which the Diamond may or may not be useful. Given the exploratory and validation-oriented nature of the interviews, the aim was not to reach theoretical saturation. Instead, the focus was on gathering meaningful insights from a select group of experienced professionals. By comparing perspectives across roles and levels of technical expertise, the study is able to surface both common themes and important contrasts. Their responses formed the core dataset for evaluating the conceptual validity and potential applicability of the AI-Fraud Diamond. 

\subsection{Comparative Summary of Interview Findings}
To provide a clear and comparative overview of the interview findings, Table 2 summarizes the perspectives of all respondents across the five key themes explored in this research. The table highlights both converging insights and nuanced differences between the interviewees, who albeit their similar audit backgrounds differ in their experiences: Pro \#1 is experienced in Fraud Risk Management, Pro \#2 in Soft Controls, Pro \#3 in Finance, and Pro \#4 in Digital Assurance. These insights are presented in more detail in subsequent subsections.

\begin{table*}[t]
\centering
\begin{tabular}{P{2cm}|P{3.3cm}|P{3.3cm}|P{3.3cm}|P{3.3cm}}
    \bf{Theme} & \bf{Consultant at Big-4 Consulting Firm A} & \bf{Senior Director at Big-4 Consulting Firm A} & \bf{Senior Associate at Big-4 Consulting Firm B} & \bf{PhD researcher in IT auditing \& ex-Auditor at Big-4 Consulting Firm B}\\
    & \bf{(Pro \#1)}& \bf{(Pro \#2)}& \bf{(Pro \#3)}& \bf{(Pro \#4)}\\
    \toprule
    \bf{Fraud in Complex Systems} & Fraud increasingly shifting toward manipulation of IT systems; traditional methods still exist & Fraud strongly linked to human behavior and organizational culture & Sampling limitations and lack of full process visibility make fraud hard to detect & System complexity obscures traceability; risks increasingly tied to cyber threats \\
    & & & &\\
    \bf{Limits of the Fraud Triangle} & Still widely used, though difficult to apply in complex scenarios & Soft controls align well with the triangle, which continues to be a valid model & Still used, but feels limited in digital or complex environments & A good starting point, but insufficient for systemic risks; needs supplementation\\ 
    & & & & \\
    \bf{Technical Opacity as a Structural Risk} & A significant and underestimated factor; limits detectability and goes beyond ‘opportunity’ & Technical opacity undermines transparency and accountability; enables fraud & Limited system insight increases risk of undetected fraud & Structural issue that limits control and enables fraud; not just technical, but also organizational \\
    & & & & \\
    \bf{Stand on the AI-Fraud Diamond} & Recognizes value of the diamond but warns of overlap with existing triangle conditions & Helpful to sharpen focus, but risks becoming too abstract & Strengthens awareness of invisible risks; however may result in a over-complicated model & Valuable in complex systems; highlights system-level risks beyond individual behavior \\
    & & & & \\
    \bf{Key Challenges in Operationalization of the Diamond} & Audit teams lack technical skills; collaboration with IT is limited and often ineffective & Effective use of the diamond requires deeper insights, clear roles, and tailored support & Increased technical knowhow within AI systems & Ability of auditors to go beyond surface-level reviews rather than accept AI system complexity as a given\\
    \bottomrule
\end{tabular}
\caption{Insights from the Interviews}
\label{table2}
\end{table*}

\subsection{Fraud in Complex Systems - Insights}

All four professionals agree that fraud is becoming increasingly difficult to detect as organizations operate in more complex, technology-driven environments. Traditional audit approaches, based on manual controls and sampling, are no longer sufficient in systems that evolve quickly and often function as black boxes. Pro \#1 explains, “\textit{Fraud is evolving more toward the manipulation of IT systems rather than just manipulating the actual figures.}” Rather than forging documents, fraudsters now exploit access rights, override controls, or alter system configurations, often without leaving clear traces.
This lack of visibility is intensified by the limited scope of audits. Pro \#3 points out, “\textit{You won’t be able to look at 100\% of the transactions}”, meaning fraud may go undetected if it falls outside the sample. Pro \#4 refers to this problem as a “\textit{technical fog}", where fraud risks are embedded in the system itself, not just in human behavior. AI tools, outsourced services, and layered platforms create situations where even internal users cannot explain how decisions are made. As Pro \#1 notes, “\textit{People don’t always really understand how programs and IT work, and those who do can take a lot of advantage.}”
Fraud is no longer only about individual actions, but also about whether systems are auditable at all. Pro \#4 asks, “\textit{Is the system even built to allow us to detect it?}” The interviews reveal a shared concern: without stronger technical insight and improved audit methods, fraud risks will remain hidden within the AI systems themselves.

\subsection{Limits of the Fraud Triangle - Insights}

The Fraud Triangle remains a widely used model due to its simplicity and behavioral focus. However, all interviewees note that in increasingly automated and opaque environments, the model faces limitations. Pro \#1 explains, “\textit{Fraud is evolving more toward the manipulation of IT systems rather than just manipulating the actual figures.}” This highlights a shift in fraud risk from individual actions to technical manipulation, where system access and settings can be exploited without leaving a visible trace.

A key challenge is how the triangle explains what creates opportunity for fraud. Traditionally, this opportunity relates to observable weaknesses, for example, within a process. However in modern systems, risk arises from invisibility rather than visibility. Pro \#4 argues, “\textit{Fraud might happen not because someone sees a gap, but because no one can see the system at all.}” Pro \#3 adds, “\textit{The structure of the AI system itself can shape the opportunity.}” These insights point to a phenomenon where fraud appears embedded in system design, making it harder to detect or even define. In such contexts, the Fraud Triangle's focus on individual behavior becomes less adequate.

Not all respondents believe the triangle needs change. Pro \#2 argues, “\textit{If a system is complex or unclear, that creates space to act unnoticed, so that’s still opportunity.}” Others, like Pro \#1, are hesitant to add new conditions without clear boundaries. Several interviewees also warn that expanding the triangle may reduce its clarity and practical value. While there is agreement that technical opacity complicates fraud detection, the core debate remains whether the triangle itself is outdated or simply needs to be interpreted through a broader lens.

\subsection{Stand on the AI-Fraud Diamond - Insights}

Interviewees are divided on whether technical opacity warrants inclusion as a separate condition. Pro \#1 and Pro \#2 believe technical opacity still fits under opportunity. “\textit{If a system is complex or unclear, that creates space to act unnoticed—so that’s still opportunity}”, says Pro \#2. Pro \#1 adds, “\textit{A lack of knowledge or technical capacity… still fits within the opportunity element.}” Both emphasize the value of maintaining the triangle’s simplicity, warning that expanding it may complicate communication and practical use.

Other professionals argue that technical opacity introduces a fundamentally different kind of risk. Pro \#3 supports the view of Pro \#4 that fraud in AI systems has largely to do with lack of system transparency. The former goes on to add: “\textit{The structure of the AI system itself can shape the opportunity.}” These insights suggest that in highly technical systems, fraud risk may be embedded in the architecture itself, rather than resulting from behavioral exploitation of a known weakness, as in the Fraud Triangle.

Several respondents highlight that technical opacity can render fraud invisible, affecting not just detection but accountability. As Pro \#1 explains, “\textit{You can influence what goes in, and you can influence what comes out, but the inside remains inaccessible.}” In such environments, the assumptions of the Fraud Triangle begin to lose relevance. However, caution remains. Pro \#2 warns that technical opacity could become a buzzword, and Pro \#3 stresses that “\textit{over-complicating the Diamond}” risks undermining its practical value. Ultimately, the question remains whether the Fraud Triangle needs to evolve, or whether auditors need better tools and interpretations to apply it within modern, opaque systems.

Table 3 summarizes the key arguments for and against adding technical opacity as a fourth condition, based solely on the views of the interviewees.

\begin{table*}[t]
\centering
\caption{Arguments for and against adding technical opacity as a fourth condition}
\begin{tabular}{p{2cm}|p{2cm}|p{12cm}}  
\textbf{Interviewee} & \textbf{Stand} & \textbf{Argument} \\
\toprule
Pro \#1 & For & Technical opacity allows fraud to happen undetected. When systems are too complex to audit, the risk goes beyond opportunity, it becomes invisible. \\

Pro \#1 & Against & Technical opacity is just a modern form of opportunity. AI systems can be exploited like any other weak process, so the Fraud Triangle already covers it. \\
&&\\
Pro \#2 & For & Naming technical opacity forces auditors to address hidden risks they would otherwise overlook. It brings blind spots into focus. \\

Pro \#2 & Against & Without clear definition, technical opacity becomes a vague label. It confuses more than it clarifies, especially for non-technical teams. \\
&&\\
Pro \#3 & For & In AI systems, fraud risk is built into the design. The system itself creates blind spots, not just the people using it. \\

Pro \#3 & Against & Adding a new layer undermines the triangle’s strength, its simplicity. More complexity makes it harder to use and less effective. \\
&&\\
Pro \#4 & For & In opaque systems, fraud hides in complexity. It can occur without clear intent or opportunity, which the current Triangle fails to capture. \\

Pro \#4 & Against & Most audit teams lack the skills to assess technical opacity. Without that expertise, adding it offers theory, not real value. \\
\bottomrule
\end{tabular}
\end{table*}

\subsection{Key Challenges in Operationalization of the Diamond - Insights}

While several interviewees support the addition of technical opacity as a fourth condition, they all agree that this conceptual step must be followed up by practical implementation. A new model alone does not improve fraud detection unless it leads to real change in how audits are performed, a concern that was raised repeatedly across interviews. Pro \#1 emphasized that many financial auditors lack the necessary technical knowledge to assess modern systems. “\textit{There’s a real need for more interdisciplinary training… many financial auditors still lack a working understanding of IT and AI systems.}” This was echoed by Pro \#3, who stressed that traditional audit skills are not enough. “\textit{What auditors need most is a mix of critical thinking and the ability to work with technical experts.}” Both professionals argue that fraud risks in opaque systems demand broader collaboration and comfort with uncertainty.

Pro \#4 added that current audit methods, such as checklists and process walkthroughs, are insufficient in automated environments. Pro \#4 called for structural changes in how audit teams operate, including integration between IT, data, legal, and finance. Pro \#4 continues to argue that auditors must be willing to go beyond surface-level reviews and “\textit{ask uncomfortable questions}”, rather than accept system complexity as a given.
Finally, Pro \#2 highlighted that even skilled auditors can only go so far without organizational support. “\textit{If conditions for proper access and cooperation aren’t met, the audit can’t continue.}” Pro \#2 also proposed that technical opacity should be broken down into specific audit questions: who controls the model? Is the logic documented? Can outcomes be verified? Such prompts help translate an abstract risk into clear, operational audit criteria.

Across all interviews, the message is consistent: the success of the AI-Fraud Diamond depends on what happens after recognition; namely, auditors must be equipped, enabled, and supported to act. Without that, technical opacity remains a theoretical risk that never reaches the audit floor.

\section{Insights, Limits, and Future Directions }

Although auditing has a crucial role to play in detecting and mitigating risks in AI systems, current approaches are not equipped to meet the challenges posed by algorithmic complexity and technical opacity. While ethical auditing frameworks exist, they lack
practical tools for identifying fraud and rarely engage with fraud directly, often overlooking the structural conditions under which AI systems can be manipulated or misused. Traditional audit strategies, such as fairness audits or performance checks, lack the precision to uncover fraudulent outputs when those outputs are buried within non-transparent models or data processes.

To address this gap, the AI-Fraud Diamond is proposed. This model expands the traditional Fraud Triangle, comprised of pressure, opportunity, and rationalization, by introducing technical opacity as a fourth condition. In contrast to behavioral models that focus on individual intent, this model accounts for the possibility that fraud may emerge not from human misconduct alone, but from system-level conditions that enable harmful or deceptive outcomes to remain undetected. It re-frames audit thinking from outcome-based detection to structural diagnosis.

The model is tested through expert interviews with audit professionals. These interviews confirmed the growing relevance of technical opacity in audit practice. Some respondents considered technical opacity to be an extension of opportunity, while others saw it as a distinct category of risk, particularly in cases where system complexity itself creates conditions of invisibility. The professionals stated key challenges in operationalizing the AI-Fraud Diamond, including limited technical expertise, inadequate interdisciplinary collaboration, and restricted access to the internal workings of AI systems. The success of the AI-Fraud Diamond depends not only on recognizing technical opacity but also on the ability of audit teams to understand and engage with it. These findings underscore the need for revised audit strategies that reflect the technical and organizational realities of modern AI systems. 

While this paper offers valuable conceptual contributions to the auditing of AI-related fraud, several limitations should be acknowledged. First, the empirical findings are based on interviews conducted with a small set of four experts, who are professionals from major audit firms. Although participants were selected for their relevant experience and technical affinity, the limited sample size may constrain the breadth of perspectives captured. The insights gathered are therefore not intended to be statistically representative, but to serve as an exploratory foundation for future research. Second, the AI-Fraud Diamond was validated conceptually through expert reflection rather than tested in practical audit settings. While the interviews provided critical feedback on the model’s relevance and usability, its operational effectiveness remains theoretical. Future research should extend this work by empirically testing the AI-Fraud Diamond in real-world audit contexts to evaluate its practical applicability and refine its theoretical foundations.

\bibliography{aaai25}

\end{document}